\documentstyle[12pt,world_sci,lnpi]{article}

\def\emline#1#2#3#4#5#6{%
       \put(#1,#2){\special{em:moveto}}%
       \put(#4,#5){\special{em:lineto}}}
\def\newpic#1{}

\title{The Electromagnetic Catalysis of the Neutrino Radiative Decay}

\author{A.A.Gvozdev, N.V.Mikheev and L.A.Vassilevskaya
\thanks{E-mail: \quad phystheo@univ.yars.free.net;
                \quad phth@cnit.yaroslavl.su}\\
{\small\it Division of Theoretical Physics, Department of Physics,}\\
{\small\it Yaroslavl State University, Yaroslavl 150000, Russia}}

\begin{document}
\begin{flushright}
{\normalsize Yaroslavl State University\\
             Preprint YARU-HE-94/04\\
             hep-ph/9409363} \\[3cm]
\end{flushright}

\maketitle

\begin{abstract}
The radiative decay of the massive neutrino
$\nu_i \rightarrow \nu_j \gamma  ( m_i > m_j )$
in the external electromagnetic field is analyzed  in  detail  in
the framework of the standard model with  lepton  mixing.  It  is
shown, that a sort of a catalysis of the neutrino radiative decay
takes place. The  effect  of  catalysis  is
significant in  the  strong electromagnetic fields ($F \gg B_e = m^2_e / e
\simeq 4.41 \cdot 10^{13}$ G) ; it is more pronounced in the case
of  the  decay  of the ultrarelativistic neutrino $(E_{\nu} \gg m_{\nu})$,
even  in  a  relatively weak electromagnetic field
($F \ll m^2_e/e$).
\end{abstract}

\vglue 3cm

\begin{center}
{\it Talk given at the VIII International Seminar "Quarks-94",\\ Vladimir,
Russia, May 11-18, 1994}
\end{center}

\newpage

\eqnumsection


\section{Introduction}

For the last two decades intensive theoretical studies  have
been given to electroweak processes with flavour violation caused
by the phenomenon of fermion  mixing.  The  description  of  this
phenomenon in  the quark sector goes back to the pioneer work  by
Cabibbo \cite{C}
and presently is put into practice  by  introducing  a
unitary $3 \times 3$ matrix $V_{ij}$ (the so  called
Cabibbo-Kobayashi-Maskawa matrix \cite{KM}.
It should be noted that qualitative progress has been
observed  in   experimental   investigation   of   quark   mixing
parameters. Except a detailed examination  of  the ``old'' mixing
angles related  to  the  first  two  quark  generations,  crucial
information  has  been  obtained  from  the  decays  of $b$-quark
containing  particles  (e.g.,  see  the  works   of   the   Argus
Collaboration on the measurement of $V_{ub}$, $V_{cb}$ \cite{FA}).
On  the  other
hand, thus far there is no experimental evidence in favor of  the
analogous mixing phenomenon in the lepton  sector.  This  can  be
accounted for in a natural way by the fact that, because  of  the
insufficiently high precision achieved in experimental studies of
neutrino-involving processes, the neutrino mass spectrum  appears
degenerate  (the  neutrinos  manifest  them  self   as   massless
particles \cite{RPP}). The  neutrino  mass  spectrum  being  degenerate,
lepton mixing is known to be purely formal and  unobservable.  At
the same time, with massive neutrino the absence of lepton mixing
seems unnatural and is virtually incompatible  with  attempts  to
somehow extend the standard model. Notice that lepton mixing,  in
itself, does not go beyond the framework of  the  standard  model
and is expected to bring forth a number of  interesting  physical
phenomena that are currently under intensive examination:

\noindent i) charged lepton radiative decays with lepton number  violation,
such as $\mu \rightarrow e \gamma$,  $\mu \rightarrow 3e$~\cite{BPP,W},
$\mu \rightarrow e \gamma \gamma$~\cite{W};

\noindent ii) neutrino radiative decays $\nu_i \rightarrow \nu_j \gamma$
(see, for example, \cite{LS}),
$\nu_i \rightarrow \nu_j \gamma \gamma$ \cite{N};

\noindent iii) neutrino oscillations \cite{BU},

\noindent iv) the possible effect of massive neutrino mixing on the
spectrum of $\beta$ -- decay-produced  electrons  (the  17  keV  neutrino
problem) \cite{L}.

Even such a short review of lepton mixing effects shows that
most of these are associated with the massive neutrino. Nowadays,
the physics of the massive  neutrino  is  becoming  a  vigorously
growing and prospective line of investigation at the junction  of
elementary particles physics, astrophysics and cosmology. It will
suffice to mention the well known problem of the  solar  neutrino
\cite{BU}
and  the  possibility  of  solving  it  (the  mechanism  of
resonance  enhancement  of  neutrino  oscillations  in  substance
\cite{MSW}),
the effect of the massive neutrino radiative decay on  the
spectrum of the relic radiation \cite{AV},
the 17 keV neutrino problem \cite{L}.
The above-mentioned beautiful  way  of  solving  the  solar
neutrino problem using the MSW mechanism \cite{MSW}
shows that the  the
massive neutrino's properties are  sensitive  to  the  medium  it
propagates  through.  Substance  is  usually  considered  as  the
medium. We note, however, that the medium can also be represented
by an  external  electromagnetic  field, which  can  significantly
influence both the properties of the massive neutrino itself \cite{ZMB}
and the process of its decay \cite{GMV}
and even  induce  novel  lepton transitions  with flavour violation
$\nu_i \leftrightarrow \nu_j$ ($i \neq j$) \cite{BTV},
forbidden in vacuum. In our preliminary communication \cite{GMV}
we have  pointed
out the probability  of  the  massive  neutrino  radiative  decay
$\nu_i \rightarrow \nu_j \gamma$ ($i \neq j$) being considerably
enhanced in  a  constant  uniform
magnetic field. Such an enhancing influence of an external  field
can be illustrated with the straightforward example of a neutrino
radiative decay in a weak (as compared with the  Schwinger  value
$F_e = m^{2}_{e}/e \simeq 4.41 \cdot 10^{13} \, G$) electromagnetic field.

To illustrate the enhancing influence of an external field  we
use the amplitude of the  Compton-like  process
$\nu_i \gamma^{\ast} \rightarrow \nu_j \gamma^{\ast}$  with
virtual photons \cite{KuM}, which, in particular, allows obtaining  the
first term of  the  expansion  of  the  radiative  decay
$\nu_i \rightarrow \nu_j \gamma$
amplitude in a weak external field. In  the  expression  for  the
amplitude of the  process
$\nu_i (p_1) + \gamma^\ast (q_1) \rightarrow  \nu_j (p_2) + \gamma^\ast(q_2)$
it is sufficient to consider $\gamma (q_2)$ as a real photon,
and to replace the field tensor of the virtual photon
$\gamma^\ast (q_1)$ by the  Fourier  image
of the external electromagnetic field tensor. Below we shall give
the expression obtained in  this  way  for  the  radiative  decay
amplitude in the  simplest  case  of  a  uniform  electromagnetic
field, in which the decay kinematics $p_1 + 0 = p_2 + q_2$ is the  same
as in vacuum. The external-field-induced contribution $\Delta {\cal M}$
to  the amplitude of the decay $\nu_i \rightarrow \nu_j \gamma$ can be
represented in the following form:

\begin{equation}
\Delta {\cal M} \simeq  {e \over 48 \pi^2} \, {G_{F} \over \sqrt{2}} \,
(jq) \, (F \tilde{f}^{*}(q)) \left < {1 \over F_{\ell}} \right > .
\label{eq:DM}
\end{equation}

\noindent where
$j_\mu = \bar \nu_j (p_2) \gamma_\mu (1 + \gamma_5) \nu_i(p_1)$,
$i,j=1,2,3$   enumerate   the
definite mass neutrino species,
$p_{1}$, $p_{2}$, $q$ are the  four-momenta  of
the initial and final neutrinos and the photon, respectively,
$F_{\mu \nu }$ is the external uniform electromagnetic field tensor,
$\tilde{f}_{\alpha \beta }(q) = \epsilon _{\alpha \beta \mu \nu}
q_{\mu} \epsilon_{\nu }(q)$,
$\epsilon _{\nu }(q)$  is  the  polarization  four-vector of the photon,
$F_\ell = m^2_\ell / e$ is the critical  value  of  the  strength
of the electromagnetic field for the charged lepton with the mass
$m_{\ell}$. We have introduced the following designation :

\begin{equation}
< A (m_\ell) > = \sum^{}_{\ell = e, \mu ,\tau} K_{i \ell} K^{*}_{j \ell}
\, A (m_\ell),    \label{eq:D<>}
\end{equation}

\noindent where  $K_{i \ell}$   ($\ell = e, \mu ,\tau$)   is   the
lepton   mixing   matrix   of Kobayashi-Maskawa type.
For the sake of comparison we  write  the
known expression for the  amplitude  of  the  neutrino  radiative
decay in vacuum \cite{LS}
which can be represented as follows:

\begin{equation}
{\cal M}_{0} \simeq  -i \;{3 e \over 32 \pi^2} \, {G_{F} \over \sqrt 2} \,
(j \tilde{f}^{*}(q) p) \left < {m^{2}_{\ell} \over m^{2}_{W}} \right > ,
\label{eq:M0}
\end{equation}

\noindent where $p = p_{1} + p_{2}$,  $m_{\ell}$, $m_{W}$ are the masses
of  the  virtual  lepton  and
$W$-boson, respectively. In  analyzing  the  amplitudes (\ref{eq:DM})
and (\ref{eq:M0})
in the case of the neutrino decay at rest, it is  necessary
to take account of $p_{1}$, $p_{2}$, $q$, $j$ being of order of the
mass  of  the decaying neutrino $m_{\nu}$.
In  this  case, the  expressions  for  the
amplitudes (\ref{eq:DM}), (\ref{eq:M0})
can be easily estimated (it is sufficient
to  allow  for  the  order  of   the   dimensional   quantities):

\begin{eqnarray}
\label{eq:ADM}
\Delta {\cal M} & \sim & G_{F} m^{3}_{\nu} ( F / F_{e}), \\
\label{eq:AM0}
{\cal M}_{0} & \sim & G_{F} m^{3}_{\nu} (m_{\tau} / m_{W})^{2}.
\end{eqnarray}

\noindent It  follows  herefrom   that,   given   the
condition

\begin{equation}
(F / F_{e})^{2} \gg (m_{\tau} / m_{W})^{4}  \label{eq:C}
\end{equation}

\noindent (here $F$ stands for the strengths of the magnetic ($B$) and
electric ($E$) fields), the probability of the decay
$\nu_i \rightarrow \nu_j \gamma$ in  an  external
field is much greater than that in vacuum, even for a  relatively
weak electromagnetic field ($10^{-3} \ll  F/F_e \ll 1$). The  catalyzing
effect of an external field becomes even more substantial in  the
case of the ultrarelativistic neutrino decay  ($E_\nu \gg m_\nu$).
With  the amplitudes (\ref{eq:DM}), (\ref{eq:M0}), being Lorentz-invariant,
the analysis can
be conveniently carried out in the rest  frame  of  the  decaying
neutrino. In this case the electromagnetic field in  Eq. (\ref{eq:ADM})
is obtained by the Lorentz  transformation  from  the  laboratory
frame, in which the external field $F$ is given, to the rest  frame
of the decaying neutrino:

\begin{equation}
F' \sim { E_\nu \over m_\nu} F \gg  F.  \label{eq:UF}
\end{equation}

\noindent Comparing the expressions (\ref{eq:ADM}) and (\ref{eq:AM0}),
in view of  (\ref{eq:UF}),   we
notice that the catalyzing effect of the external  field  becomes
appreciable under a much weaker condition, as compared to (\ref{eq:C}):

\begin{equation}
{(p_1 F F p_1) \over m^2_\nu F^2_e} \gg
\left ( {m_{\tau} \over m_{W}} \right )^4.    \label{eq:UE}
\end{equation}

In  this  case  the  ratio  between  the  probabilities  of   the
ultrarelativistic  neutrino  decay $w^{(F)}$   and   the   decay   in
vacuum $w^{(0)}$  is  of  the  order:

\begin{equation}
{w^{(F)} \over w^{(0)}} \sim
\left ( {F \over F_{e}} \right )^2
\left ( {E_{\nu } \over m_{\nu}} \right )^2
\left ( {m_{W} \over m_{\tau}} \right )^4 \gg 1.  \label{eq:RW}
\end{equation}

\noindent The expression (\ref{eq:RW}) shows that in the ultrarelativistic
neutrino decay the enhancement is mainly due to a decrease  in  the  decay
probability suppression by the smallness  of  the  neutrino  mass
($w^{(F)} \sim m^{4}_{\nu}$, $w^{(0)} \sim m^{5}_{\nu} (m_{\nu} / E_{\nu})$).
It is  natural  to  expect  that  in
taking the account of further  terms  in  the  expansion  of  the
amplitude of the radiative  decay $\nu_i \rightarrow \nu_j \gamma$
with  respect  to  the
external field, the suppression  mentioned  above  can  be  fully
canceled.  All  this  makes  it  interesting  to  calculate   the
amplitude with the  external  electromagnetic  field  taken  into
account exactly. An expression thus obtained will be valid in the
case of the  neutrino  radiative  decay $\nu_i \rightarrow \nu_j \gamma$
in  an  external
electromagnetic field, which has not to be weak  as  against  the
Schwinger value $F_e$.

\section{The crossed fields}

At  present the  experimentally  accessible  strengths  of
electromagnetic  fields  are  significantly  below  the  critical
strength ($F/F_e \ll  1$, $F=B,{\cal E}$,
$F_{e}= m^{2}_{e}/e \simeq 4.41 \cdot 10^{13} \, G$). Because  of
this  ,  field-induced  effects  are  especially  marked  in  the
ultrarelativistic  case  with  the   dynamic   parameter

\begin{equation}
\chi ^{2} = {e^{2}(pFFp) \over m^{6}}   \label{eq:DP}
\end{equation}

\noindent being not small even for  a  relatively  weak  field
($F_{\mu \nu }$ is  the
external field tensor,  $p_{\alpha }$ is the 4-momentum,
$m$ is the mass of the
particle). This is due to  the  fact  that  in  the  relativistic
particle rest frame the field  may  turn  out  of  order  of  the
critical one or even higher, appearing very close to the constant
crossed fields.  Thus, the calculation  in constant crossed fields
($\vec {\cal E} \perp \vec B$, ${\cal E}=B$)  is  relativistic  limit
of  the  calculation  in  an
arbitrary  weak  smooth  field,  possesses  a  great  extent   of
generality and acquires interest by  itself.  We  note  that,  as
($FF=F\tilde{F}=0$)  in crossed fields, the dynamic parameter
$\chi ^{2}$  (\ref{eq:DP})  is
the single field invariant, by which  the  decay  probability  is
expressed.


\unitlength=0.80mm
\special{em:linewidth 0.4pt}
\linethickness{0.4pt}
\begin{picture}(141.50,65.50)
\put(21.00,39.05){\oval(8.00,3.00)[l]}
\put(21.00,36.55){\oval(4.00,2.00)[rt]}
\put(25.00,36.55){\oval(4.00,2.00)[lb]}
\put(25.00,34.55){\oval(4.00,2.00)[rt]}
\put(29.00,34.55){\oval(4.00,2.00)[lb]}
\put(29.00,32.55){\oval(4.00,2.00)[rt]}
\put(33.50,32.55){\oval(5.00,2.00)[b]}
\put(38.00,32.55){\oval(4.00,2.00)[lt]}
\put(38.00,34.55){\oval(4.00,2.00)[rb]}
\put(42.00,34.55){\oval(4.00,2.00)[lt]}
\put(42.00,36.55){\oval(4.00,2.00)[rb]}
\put(46.00,36.55){\oval(4.00,2.00)[lt]}
\emline{45.00}{37.55}{1}{47.00}{37.53}{2}
\emline{21.00}{40.55}{3}{47.00}{40.54}{4}
\put(48.00,39.05){\oval(6.00,3.00)[r]}
\emline{47.00}{40.55}{5}{48.00}{40.54}{6}
\emline{46.00}{37.55}{7}{49.00}{37.53}{8}
\put(21.50,39.05){\oval(7.00,1.00)[l]}
\put(22.00,37.55){\oval(4.00,2.00)[rt]}
\put(26.00,37.55){\oval(4.00,2.00)[lb]}
\put(26.00,35.55){\oval(4.00,2.00)[rt]}
\put(30.00,35.55){\oval(4.00,2.00)[lb]}
\put(30.00,33.55){\oval(4.00,2.00)[rt]}
\put(33.50,33.55){\oval(3.00,2.00)[b]}
\put(37.00,33.55){\oval(4.00,2.00)[lt]}
\put(37.00,35.55){\oval(4.00,2.00)[rb]}
\put(41.00,35.55){\oval(4.00,2.00)[lt]}
\put(41.00,37.55){\oval(4.00,2.00)[rb]}
\put(45.00,37.55){\oval(4.00,2.00)[lt]}
\put(46.50,39.05){\oval(7.00,1.00)[r]}
\emline{21.00}{39.55}{9}{47.00}{39.55}{10}
\emline{44.00}{38.55}{11}{47.00}{38.55}{12}
\put(1.00,39.55){\vector(1,0){10.00}}
\put(51.00,39.55){\vector(1,0){10.00}}
\emline{11.00}{39.55}{13}{17.00}{39.55}{14}
\emline{61.00}{39.55}{15}{68.00}{39.55}{16}
\put(94.33,34.28){\oval(8.00,3.00)[l]}
\put(94.33,36.78){\oval(4.00,2.00)[rb]}
\put(98.33,36.78){\oval(4.00,2.00)[lt]}
\put(98.33,38.78){\oval(4.00,2.00)[rb]}
\put(102.33,38.78){\oval(4.00,2.00)[lt]}
\put(102.33,40.78){\oval(4.00,2.00)[rb]}
\put(106.83,40.78){\oval(5.00,2.00)[t]}
\put(111.33,40.78){\oval(4.00,2.00)[lb]}
\put(111.33,38.78){\oval(4.00,2.00)[rt]}
\put(115.33,38.78){\oval(4.00,2.00)[lb]}
\put(115.33,36.78){\oval(4.00,2.00)[rt]}
\put(119.33,36.78){\oval(4.00,2.00)[lb]}
\emline{118.33}{35.78}{17}{120.33}{35.78}{18}
\emline{94.33}{32.78}{19}{120.33}{32.78}{20}
\put(121.33,34.28){\oval(6.00,3.00)[r]}
\emline{120.33}{32.78}{21}{121.33}{32.78}{22}
\emline{119.33}{35.78}{23}{122.33}{35.78}{24}
\put(94.83,34.28){\oval(7.00,1.00)[l]}
\put(95.33,35.78){\oval(4.00,2.00)[rb]}
\put(99.33,35.78){\oval(4.00,2.00)[lt]}
\put(99.33,37.78){\oval(4.00,2.00)[rb]}
\put(103.33,37.78){\oval(4.00,2.00)[lt]}
\put(103.33,39.78){\oval(4.00,2.00)[rb]}
\put(106.83,39.78){\oval(3.00,2.00)[t]}
\put(110.33,39.78){\oval(4.00,2.00)[lb]}
\put(110.33,37.78){\oval(4.00,2.00)[rt]}
\put(114.33,37.78){\oval(4.00,2.00)[lb]}
\put(114.33,35.78){\oval(4.00,2.00)[rt]}
\put(118.33,35.78){\oval(4.00,2.00)[lb]}
\put(119.83,34.28){\oval(7.00,1.00)[r]}
\emline{94.33}{33.78}{25}{120.33}{33.78}{26}
\emline{117.33}{34.78}{27}{120.33}{34.78}{28}
\put(74.33,33.78){\vector(1,0){10.00}}
\put(124.33,33.78){\vector(1,0){10.00}}
\emline{84.33}{33.78}{29}{90.33}{33.78}{30}
\emline{134.33}{33.78}{31}{141.33}{33.78}{32}
\put(34.00,40.32){\circle*{1.72}}
\put(107.00,41.39){\circle*{1.72}}
\put(34.00,41.83){\oval(2.00,2.15)[r]}
\put(34.00,43.77){\oval(2.00,1.72)[l]}
\put(34.17,45.70){\oval(1.67,2.15)[r]}
\put(34.17,47.64){\oval(1.67,1.72)[l]}
\put(34.17,49.57){\oval(1.67,2.15)[r]}
\put(34.17,51.51){\oval(1.67,1.72)[l]}
\put(34.17,53.23){\oval(1.67,1.72)[r]}
\put(107.00,43.33){\oval(2.00,2.15)[r]}
\put(107.16,45.26){\oval(1.67,1.72)[l]}
\put(107.16,47.20){\oval(1.67,2.15)[r]}
\put(107.33,49.13){\oval(1.33,1.72)[l]}
\put(107.16,51.07){\oval(1.67,2.15)[r]}
\put(107.33,53.01){\oval(2.00,1.72)[l]}
\put(107.50,54.73){\oval(1.67,1.72)[r]}
\emline{26.00}{42.05}{33}{28.00}{39.89}{34}
\emline{28.00}{39.89}{35}{26.00}{38.17}{36}
\emline{38.33}{42.05}{37}{40.33}{39.89}{38}
\emline{40.33}{39.89}{39}{38.33}{38.17}{40}
\emline{105.33}{34.51}{41}{107.66}{33.22}{42}
\emline{107.66}{33.22}{43}{105.33}{31.50}{44}
\put(38.00,52.55){\makebox(0,0)[cc]{$\gamma$}}
\put(111.33,52.78){\makebox(0,0)[cc]{$\gamma$}}
\put(11.00,43.55){\makebox(0,0)[cc]{$\nu_i$}}
\put(63.00,43.55){\makebox(0,0)[cc]{$\nu_j$}}
\put(25.00,45.05){\makebox(0,0)[cc]{$l^-$}}
\put(40.00,45.05){\makebox(0,0)[cc]{$l^-$}}
\put(34.00,27.05){\makebox(0,0)[cc]{\small{W} , $\varphi$ }}
\put(84.33,37.28){\makebox(0,0)[cc]{$\nu_i$}}
\put(136.33,37.28){\makebox(0,0)[cc]{$\nu_j$}}
\put(117.33,43.28){\makebox(0,0)[cc]{\small{W} , $\varphi$}}
\put(95.33,43.28){\makebox(0,0)[cc]{\small{W} , $\varphi$}}
\put(107.33,26.78){\makebox(0,0)[cc]{$l^-$}}
\put(34.00,18.50){\makebox(0,0)[cc]{\big{(}\large{a}\big{)}}}
\put(106.33,18.28){\makebox(0,0)[cc]{\big{(}\large{b}\big{)}}}
\put(70.00,8.00){\makebox(0,0)[cc]{\large Fig.~1}}
\end{picture}


In the lowest order of the  perturbation  theory,  a  matrix
element of the radiative decay of  the  massive  neutrino
$\nu _{i} \rightarrow \nu _{j}\gamma $  ($i \neq j$)
in Feynman gauge is described by the diagrams,  represented
in Fig.1, where double lines imply the influence of the  external
field in the propagators of $W$- and $\phi$-bosons and charged
leptons ($\ell = e, \mu, \tau$).
Under  the
conditions $m^{2}_{\ell }/m^{2}_{W}\ll  1$,  $e F / m^{2}_{W} \ll 1$
the     field     induced
contribution $\Delta {\cal M}^{(F)}= {\cal M}-{\cal M}^{(0)}$  to  the
decay  amplitude  can   be
calculated in the local limit, in which the lines $W$  and $\varphi $  are
contacted to a point, as it is shown in Fig.2.
It is most easily seen,  if $\Delta {\cal M}^{(F)}$
is expanded to a series in terms of the  external  field.
We note that with  the  orthogonality  of  the
mixing matrix $K_{ij}$  taken  into  account the  main contribution
to  the  integral
over momentum in the  loop  gives  from  the  region  of  the
virtual momenta $p \sim m_{\ell } \ll  m_{W}$.  We  remind  that  we  are
investigating flavor violating processes ($i\neq j$), and, hence, $<A>=0$,
if $A$ is independent of $m_{\ell }$ . Thus, the dominant  contribution  of
order $ 1 / m^{2}_{W} \sim G_{F}$ only comes from the diagrams with
one $W$-propagator in Fig.1.

Even such a simple analysis shows the following:

\noindent i) of the diagrams in Fig.1, the predominant contribution  is
made by the diagram ($a$), which, in the  local  limit  of $W$-boson
propagator being contracted to a point, transforms to the diagram
shown in Fig.~2;

\vspace{\baselineskip}


\unitlength=0.75mm
\special{em:linewidth 0.4pt}
\linethickness{0.4pt}

\begin{center}
\begin{picture}(50.00,57.00)
\put(10.00,30.00){\vector(1,0){10.00}}
\put(20.00,30.00){\vector(1,0){20.00}}
\emline{40.00}{30.00}{1}{50.00}{30.00}{2}
\put(30.00,38.00){\circle{16.00}}
\put(30.00,38.00){\circle{13.00}}
\put(30.00,31.00){\circle*{2.50}}
\put(35.00,52.00){\makebox(0,0)[cc]{$\gamma$}}
\put(42.00,37.00){\makebox(0,0)[cc]{\cal f}}
\put(18.00,37.00){\makebox(0,0)[cc]{\cal f}}
\put(18.00,25.00){\makebox(0,0)[cc]{$\nu_i$}}
\put(42.00,25.00){\makebox(0,0)[cc]{$\nu_i$}}
\put(30.00,15.00){\makebox(0,0)[cc]{\large Fig.~2.}}
\put(30.00,46.50){\oval(3.00,3.00)[r]}
\put(30.00,49.50){\oval(3.00,3.00)[l]}
\put(30.00,52.50){\oval(3.00,3.00)[r]}
\put(30.00,55.50){\oval(3.00,3.00)[l]}
\put(30.00,45.00){\circle*{2.50}}
\end{picture}
\end{center}


\vspace{-0.5\baselineskip}

\noindent ii) since in calculating $\Delta {\cal M}$ the  mass  of
the $W$-boson  in  the
local limit only appears in the  weak  interaction  constant
$G _{F}=g^{2}/8m^{2}_{W}$,   the  amplitude  does  not  contain
the   known   GIM suppression
factor of the decay $\nu _{i}\rightarrow \nu _{j}\gamma $ in vacuum
$\sim m^{2}_{\ell } / m^{2}_{W} \ll  1$ (see  Eq.(\ref{eq:M0})).
The expression for the  amplitude,  corresponding  to  the
diagram in Fig.~2, can be represented in the following form:

\begin{eqnarray}
\label{eq:Am1}
\Delta {\cal M} & = & {ieG_{F}\over \sqrt 2} \; j_{\beta }
\epsilon ^{*}_{\alpha }(q) <J_{\alpha \beta }(q)> - {\cal M}^{(0)}, \\
\label{eq:Int1}
J_{\alpha \beta }(q) & = & \int d^{4}X Tr\left[ \gamma _{\alpha } \hat{S}(X)
\gamma _{\beta } (1+\gamma _{5}) \hat{S}(-X) \right] e^{iqX},
\end{eqnarray}

\noindent where $X=x-y$ and $\hat{S}(X)$ is the propagator of a
charged  lepton  in  the
crossed fields (see Appendix A, Eqs.(A.1) and (A.2)).
All the other quantities in
(\ref{eq:Am1}) are defined above (see Eqs.(\ref{eq:DM})-(\ref{eq:M0})).
The details of the
tensor $J_{\alpha \beta }(q)$ calculation may be found in Appendix A,
while here we only give the result of the calculation:

\begin{eqnarray}
\label{eq:Am2}
\Delta {\cal M} & = & {e G_{F} \over 4 \pi^2 \sqrt 2}
\bigg < e (\tilde{F}f^{*}) \,
{(qFFj) \over (qFFq)} \, I_1 + {e \over 8 m^2_\ell} \,
(F \tilde{f}^{*})(qj) \, I_{2}  \nonumber \\
& + & {e^2 \over 24 m^4_\ell} \, (F \tilde{f}^{*}) \, (q\tilde{F}j) \,
I_3 + {e^2 \over 48 m^4_\ell} \, (Ff^{*}) \, (qFj) \, I^4 \bigg >, \\
f_{\alpha \beta} & = & q_\alpha \epsilon_\beta - q_\beta \epsilon_\alpha,
\nonumber \\
\tilde{f}_{\mu \nu } & = & {1 \over 2} \epsilon_{\mu \nu \alpha \beta} \,
f_{\alpha \beta}, \nonumber
\end{eqnarray}

\noindent where $F_{\mu \nu}$,
$\tilde{F}_{\mu \nu} = \epsilon_{\mu \nu \alpha \beta} \,
F_{\alpha \beta} / 2$
are  the  constant  field  and  dual
tensors; $e > 0$  is  the  elementary  charge,  $G_F$  is  the  Fermi
constant. In Eq.(\ref{eq:Am2}) $I_a$ ($a = 1, \ldots, 4$)
are integrals  of  the  known Hardi-Stoks functions $f(u)$:

\begin{eqnarray}
I_1 & = & \int^{1}_{0} dt \, uf(u),  \nonumber \\
\label{eq:Int2}
I_2 & = & \int^{1}_{0} dt \, (1 - t^2) \, uf(u), \\
I_3 & = & \int^{1}_{0} dt \, (1 - t^2) \, (3 - t^2) \, u^2 \,
{df \over du}, \nonumber \\
I_4 & = & \int^{1}_{0} dt \, (1 - t^2) \, (3 + t^2) \, u^2 \,
{df \over du}, \nonumber \\
\label{eq:HS}
f(u) & = & i \; \int^{\infty }_{0} dz \exp \left [ -i \,
(zu + {1 \over 3} z^3) \right ], \\*[.5\baselineskip]
u & = &
( e^2 (qFFq) / 16 m^6_\ell )^{- 1/3} \, (1 - t^2)^{- 2/3} \nonumber
\end{eqnarray}

\noindent As can be readily checked, the amplitude (\ref{eq:Am2})
is evidently gauge invariant,
as it is expressed in terms  of  the  tensors  of  the
external field $F_{\mu \nu}$ and the photon field $f_{\mu \nu}$.
In the  weak  field limit
($F/F_e \ll 1$) the predominant contribution to the amplitude  is
made by the second term in the eåpression (\ref{eq:Am2}),
as the integrals $I_a$ can be easily evaluated in  this limit,
taking  into  account
the orthogonality of the lepton mixing matrix $K_{i \ell}$ :

\begin{eqnarray}
< I_1 > & = & < 1 + O [(F/F_e)^2] > \; \sim \; O [(F/F_e)^2], \nonumber \\
\label{eq:EInt2}
I_2 & \simeq & 2 / 3 , \\
I_3 & \simeq & 28 / 15 , \nonumber \\
I_4 & \sim & O [(F/F_e)^4].    \nonumber
\end{eqnarray}

\noindent As expected, the amplitude (\ref{eq:Am2})
in view of (\ref{eq:EInt2})
in the weak field limit coincides with the expression (\ref{eq:DM}).
The amplitude  of  the process $\nu_i \rightarrow \nu_j \gamma$
in  a  crossed  fields  (\ref{eq:Am2})
is   substantially simplified in two cases:
that of the decay of a neutrino at  rest ($E_\nu = m_\nu$) and
that of the decay of  an  ultrarelativistic  neutrino ($E_\nu \gg m_\nu$).

\subsection{Neutrino at rest ($E_\nu = m_\nu$)}

In this case the dynamic parameter (\ref{eq:DP})

\begin{displaymath}
\chi^2_\ell = {e^2 (p_1 F F p_1) \over m^6_\ell} =
\left ( {m_\nu \over m_\ell} \; {e F \over m^2_\ell} \right )^2
\end{displaymath}

\noindent is obviously small even for the fields, the  strengths  of  which
are of order or greater than the critical one ($F \ge  m^2_\ell / e$,
$m_\nu \ll m_\ell$, $\chi_\ell \ll  1$),
and the decay amplitude (\ref{eq:Am2}), (\ref{eq:Int2}) is:

\begin{eqnarray}
\Delta {\cal M} & \simeq & {e G_{F} \over 60 \pi^2 \sqrt 2} \,
\bigg \lbrace ({\cal F} \tilde{f}^*) \left [ (j {\cal F} {\cal F} q) +
{5 \over 4} (jq) - {7 \over 6} (q \tilde{\cal F} j) \right ] \nonumber \\
& - & {19 \over 24} ({\cal F} f^*) (q {\cal F} j) \bigg \rbrace
(K_{ie} K^*_{je}).   \label{eq:Am3}
\end{eqnarray}

Here we have introduced the field tensor
${\cal F}_{\mu \nu} = F_{\mu \nu} / F_e$,
made dimensionless  by  the  critical  value $F_e = m^2_e / e$,
$m_e$  is   the electron  mass.
It  is  clear  from Eq.(\ref{eq:Am3})  that  the  decay
probability is represented by a polynomial  of  sixth  degree  in
field strength. In  the  limit $F \ll  F_e$ (weak  field  limit)  the
expression for the decay probability is governed  by  the  lowest
power of $F$  and has the following form:

\begin{equation}
w_{weak} \simeq  {\alpha \over 18 \pi } \; {G^2_F \over 192 \pi^3} \;
m^5_i \, \left ( 1 - {m^2_j \over m^2_i} \right )^5 \,
\left ( {F \over F_e} \right )^2 \,
| K_{ie} K^*_{je} |^2,      \label{eq:Ww}
\end{equation}

\noindent In the opposite case $F \gg F_e$ (strong field limit) we have:

\begin{equation}
w_{st} \simeq  {\alpha \over 4 \pi} \; {G^2_F \over (15 \pi)^3} \;
m^5_i \, \left ( 1 - {m^2_j \over m^2_i} \right ) \,
\left ( 1 + 5 {m^2_j \over m^2_i} \right ) \,
\left ( {F \over F_e} \right )^6
| K_{ie} K^*_{je} |^2.       \label{eq:Ws}
\end{equation}

Here $m_i$ is the mass of the initial neutrino,  $m_j$  is  that  of
the final one. This expression should be compared with  the  well
known probability of the decay $\nu_i \rightarrow \nu_j \gamma$ in vacuum:

\begin{equation}
w_0 \simeq  {27 \alpha \over 32 \pi} \; {G^2_F \over 192 \pi^3} \;
m^5_i \, \left ( {m_\tau \over m_W} \right )^4 \,
\left ( 1 + {m^2_j \over m^2_i} \right ) \,
\left ( 1 - {m^2_j \over m^2_i} \right )^3 \,
| K_{i\tau} K^*_{j\tau} |^2.    \label{eq:W0}
\end{equation}

\noindent The comparison shows that the enhancement of the probability of
the decay in a crossed field does take  place,  as  there  is  no
suppression $\sim (m^2_\ell / m^2_W)$ in Eqs.(\ref{eq:Ww}) and (\ref{eq:Ws}).
Besides, the decay in strong crossed fields (\ref{eq:Ws}) is catalyzed
by another factor $\sim (F / F_e)^6 \gg 1$.

\subsection{Ultrarelativistic neutrino ($E_\nu \gg m_\nu $)}

Notice that in the ultrarelativistic limit the kinematics  of
the decay $\nu_i (p_1) \rightarrow \nu_j (p_2) + \gamma (q)$ is such
that the momentum  4-vectors of the initial neutrino $p_1$
and the decay products $p_2$  and $q$  are
almost parallel to each other. Therefore, the current 4-vector
$j_\alpha = \bar \nu_j (p_2) \gamma_\alpha (1 + \gamma_5) \nu_i (p_1)$
is also proportional to these vectors
$(j_\alpha \sim p_{1 \alpha} \sim q_\alpha \sim p_{2 \alpha}$).
The amplitude expression (\ref{eq:Am2}) in  this  case
is significantly reduced to give

\begin{equation}
\Delta {\cal M} \simeq  {e^2 G_F \over \pi^2} \;
(\epsilon^* \tilde F p_1) \;
\left [ (1-x) + {m^2_j \over m^2_i} (1+x)
\right ]^{\mbox{\normalsize $1 \over 2$}} \;
< I_1 >,      \label{eq:Am4}
\end{equation}

\noindent where $x=\cos \vartheta$, $\vartheta$ is the angle between
the vectors $\vec p_1$ (the momentum of the decaying ultrarelativistic
neutrino) and $\vec {q'}$ (the photon  momentum  in  the  decaying
neutrino $\nu_i$ rest frame). The argument $u$ of the Hardi-Stoks
function $f(u)$  in  the integral $I_1$ (see Eq.(\ref{eq:Int2}))
in the ultrarelativistic case has the form:

\begin{equation}
u = 4 \left [ (1+x) (1-t^2)
\left ( 1 - {m^2_j \over m^2_i} \right ) \,
\chi_\ell \right ]^{\mbox{\normalsize - ${2 \over 3}$}}.  \label{eq:Uu}
\end{equation}

The Lorentz-invariant decay probability $w E_\nu$ can therewith be
expressed as an integral of the amplitude squared with respect to the
variable $x$:

\begin{eqnarray}
w E_\nu & \simeq & {1 \over 16 \pi} \,
\left ( 1 - {m^2_j \over m^2_i} \right ) \;
\int\limits^{+1}_{-1} dx \; | \Delta {\cal M} |^2  \\
& = & {\alpha \over 4 \pi} \; {G^2_F \over \pi^3} m^6_e \chi^2_e \,
\left ( 1 - {m^2_j \over m^2_i} \right )
\int\limits^{+1}_{-1} dx \left [ (1-x) + {m^2_j \over m^2_i} (1+x) \right ] \,
| < I_1> |^2. \nonumber       \label{eq:WEe}
\end{eqnarray}

\noindent For small values of the dynamic parameter ($\chi_\ell \ll 1$),
the integral $I_1 (\chi_\ell)$ is expanded into the following series:

\begin{eqnarray}
I_1 & \simeq & 1 + {1 \over 15} \tilde{\chi}^2_\ell +
{4 \over 63} \tilde{\chi}^4_\ell + \ldots,  \nonumber \\
\tilde{\chi}_\ell & = & {1+x \over 2} \;
\left ( 1 - {m^2_j \over m^2_i} \right ) \; \chi_\ell,
\label{eq:Int3}
\end{eqnarray}

\noindent and the probability (\ref{eq:WEe}) can be represented in the form:

\begin{equation}
wE_\nu \simeq {\alpha \over 4 \pi} \; {G^2_F \over (15 \pi)^3} \;
m^6_e \chi^6_e \;
\left ( 1 - {m^2_j \over m^2_i} \right )  \,
\left ( 1 + 5 {m^2_j \over m^2_i} \right )  \,
| K_{ie} K^*_{je} |^2.   \label{eq:WEs}
\end{equation}

\noindent For great values of $\chi_\ell \gg 1$, using the asymptotic
behavior of the Hardi-Stoks function both at great and at small values
of the argument and also the unitarity of the mixing matrix $K_{i \ell}$,
one can represent Eq.(\ref{eq:WEe}) in the form:

\begin{eqnarray}
\label{eq:WEL1}
wE_\nu & \simeq & {\alpha \over 4 \pi} \; {G^2_F \over \pi^3} \;
m^6_e \chi^2_e \; \left ( 1 - {m^4_j \over m^4_i} \right ) \;
\Bigg \lbrace
\begin{array}{l}
| K_{ie} K^*_{je} |^2, \; \chi_e \gg 1, \; \chi_{\mu, \tau} \ll 1, \\[1mm]
| K_{i \tau} K^*_{j \tau} |^2, \; \chi_e \gg \chi_\mu \gg 1, \;
\chi_\tau \ll 1,
\end{array}  \\
\label{eq:WEL2}
wE_\nu & \simeq & {21.7 \alpha \over \pi} \; {G^2_F \over \pi^3} \;
m^6_\tau \chi_\tau \; | K_{i \tau} K^*_{j \tau} |^2,  \\
& & \chi_\tau \, \left ( 1- {m^2_j \over m^2_i} \right ) \gg  1.
\nonumber
\end{eqnarray}

\noindent As the dynamic parameter
$\chi_\ell \sim (E_\nu / m_\ell) (F / F_\ell)$ is proportional to the
neutrino's energy, it is clear from Eqs.(\ref{eq:WEs})--(\ref{eq:WEL2})
that, with increasing the energy of the decaying neutrino, the decay
probability scales first as $\sim E^5_\nu$ ($\chi_e \ll 1$),
then linearly $\sim E_\nu$ ($\chi_e \gg 1$, $\chi_\tau \ll 1$)
and, at last, becomes constant ($\chi_\tau \gg 1$).
Comparing the expression of the cross  field  decay  probability
(\ref{eq:WEs})--(\ref{eq:WEL2})
with the vacuum decay  probability  (\ref{eq:W0}), we notice that the
catalyzing effect of the field on the ultrarelativistic  neutrino
decay becomes even more substantial  compared  to  the  situation
with the neutrino at rest,  because  there  is  no  suppression
caused by the smallness of the mass of the decaying neutrino.
Recall   also   that   in   all eåpressions   of  the
cross-field  decay   probability   the   well-known   suppression
GIM-factor $(m_\ell / m_W)^4$  characteristic of the  probability
of the decay $\nu_i \rightarrow \nu_j \gamma$ in  vacuum  is  lacking,
whereas the decay of a neutrino at rest in a strong crossed field
(see Eq.(\ref{eq:Ws})) is accelerated by an extra  factor
$\sim (F / F_e)^6 \gg 1$.
The lack of suppression by the smallness of the neutrino mass  in
the  decay probability in the ultrarelativistic case
(\ref{eq:WEs})--(\ref{eq:WEL2}) may be useful in discussing prospective
laboratory experiments aimed  at
discovery of the radiative decay of  neutrinos  from  high-energy
accelerators in an almost crossed external electromagnetic  field
(for example, in a monocrystal's field).

\appendixes

\vspace{\baselineskip}

\section{$J_{\alpha \beta}$ calculation
in constant electromagnetic field}

\vspace{0.5\baselineskip}

The amplitude corresponding  to  the  diagram  in  Fig.~3  is
calculated according to the conventional Feinman rules. In  doing
so, for the propagators of intermediate charged leptons and quarks
exact solutions are used of the corresponding wave  equations  in
the constant electromagnetic field. With the crossed fields,  the
propagator of the charged fermion $\hat S^{(F)} (x, y)$
in the proper time formalism~\cite{S} has the form:

\begin{equation}
\label{eq:SFA}
\hat S^{(F)} (x, y) = e ^{\mbox{\normalsize $i \Phi (x,y)$}}
\hat S (X) ,
\end{equation}
\begin{eqnarray}
\label{eq:S1}
\hat S (X) & = & - {i \over 16 \pi^2} \int\limits_0^\infty
{ds \over s^2} \bigg [ {1 \over 2s} (X \gamma) -
{i e Q_f \over 2} (X \tilde F \gamma) \gamma_5 -
{s e^2 Q_f^2 \over 3} (X F F \gamma) + m
\nonumber \\
& + & {s m e Q_f \over 2} (\gamma F \gamma) \bigg ]
\exp \bigg [ -i \left ( s m^2 + {1 \over 4 s} X^2 +
{s e^2 Q_f^2 \over 12} (XFFX) \right ) \bigg ] ,
\end{eqnarray}

\noindent where $X_\mu = (x-y)_\mu$, $F_{\mu \nu}$, $\tilde F_{\mu \nu}$
are the field tensor  and  field  dual tensor, $Q_f$ is the relative
fermion charge,
$e > 0$ is  the  elementary  charge, $\gamma_\mu$, $\gamma_5$
are  Dirac $\gamma$-matrices (the metric, the convensional representation
of  Dirac $\gamma$-matrices, etc. correspond to the book~\cite{BLP}),
$m$ is the mass of the charge fermion, the phase $\Phi (x,y)$ is determined
in  the following way:

\begin{eqnarray}
\Phi (x,y) & = & - e Q_f \,
\int\limits^{\mbox{\normalsize $x$}}_{\mbox{\normalsize $y$}}
d\xi_\mu \, K_\mu (\xi),
\nonumber \\
K_\mu (\xi) & = & A_\mu (\xi) + {1 \over 2} F_{\mu \nu} (\xi - y)_\nu.
\label{eq:FKA}
\end{eqnarray}

\noindent Owing to $\partial_\mu K_\nu - \partial_\nu K_\mu = 0$,
the path of integration from $y$ to $x$ in~(\ref{eq:FKA})
is arbitrary. Using

\begin{displaymath}
\Phi (x,y) + \Phi(y,x) = 0 ,
\end{displaymath}

\noindent the integration of $J_{\alpha \beta}$
with respect to $x$ and $y$ can be easily redused to an
integration  with  respect  to $X = x - y$ (see Eq.~(\ref{eq:Int1})).
{}From~(\ref{eq:Int1}) and~(\ref{eq:S1}) it is  clear  that
the integrals with respect to $X$ are Gaussian, so that they can  be
readily calculated:

\begin{eqnarray}
G & = & \int d^4X \,
e^{\mbox{\normalsize $- i \left ( \frac{1}{4} XRX + qX \right )$}}
= - (4 \pi )^2 (\det R)^{- 1/2}
e^{\mbox{\normalsize $(i q R^{-1} q)$}},
\nonumber \\
G_\mu & = & \int d^4X \, X_\mu \,
e^{\mbox{\normalsize $- i \left ( {1 \over 4} XRX + qX \right )$}}
= i {\partial G \over \partial q^\mu},   \\
G_{\mu \nu} & = & \int d^4X \, X_\mu X_\nu
e^{\mbox{\normalsize $- i \left ( {1 \over 4} XRX + qX \right )$}}
= - {\partial^2 G \over \partial q^\mu \partial q^\nu}.
\nonumber
\label{eq:SG}
\end{eqnarray}

\noindent In the remaining double integral with respect to the proper times
$s_1$, $s_2$, it is convenient to pass to  the  dimensionless  variables
$z$, $t$:

\begin{eqnarray}
z = m^2 (s_1 + s_2), & & t = {s_1 - s_2 \over s_1 + s_2}, \qquad
ds_1 ds_2 = {1 \over 2 m^4} z dz dt,
\nonumber \\
0 \le z \le \infty, & & -1 \le t \le 1,
\label{eq:NV}
\end{eqnarray}

\noindent The substitution into the amplitude~(\ref{eq:Am1}) of the
expression  for $J_{\alpha \beta}$ in the form of a double
integral with respect to $z$, $t$ results
in the final expression~(\ref{eq:Am2}) and~(\ref{eq:Int2}).

\vspace{1.5\baselineskip}

\noindent {\bf Acknowledgements}

\vspace{0.5\baselineskip}

\noindent The work presented here was supported in part by
Grant N RO 3000 from the International Science Foundation.
The organizers of the QUARKS-94 Symposium deserve a special
acknowledgement for organizing a very successful conference
under difficult circumstances.


\begin{thebibliography}{20}
\bibitem{C}
    N.~Cabibbo, {\it Phys. Rev. Lett.} {\bf 10} (1963) 531.
\bibitem{KM}
    M.~Kobayashi and T.~Maskawa, {\it Progr. Theor. Phys.} {\bf 49} (1973) 652.
\bibitem{FA}
    R.~Fulton et al., {\it Phys. Rev. Lett.} {\bf 64} (1990) 16;
    H.~Albrecht et al., {\it Phys. Lett.} {\bf B255} (1991) 297.
\bibitem{RPP}
    K.~Hikasa et al., {\it Review of Particle Properties,  Phys. Rev.}
    {\bf D45}, VI.1 (1992).
\bibitem{BPP}
    S.M.~Bilenky, S.T.~Petcov and B.~Pontecorvo,  {\it Phys. Lett.}
    {\bf B67} (1977) 309.
\bibitem{W}
    H.K.~Walter, {\it Nucl. Phys.} {\bf B279} (1987) 133.
\bibitem{LS}
    B.W.~Lee and R.E.~Shrock, {\it Phys. Rev.} {\bf D16} (1977) 1444.
\bibitem{N}
    J.F.~Nieves, {\it Phys. Rev.} {\bf D28} (1983) 1664.
\bibitem{BU}
    J.N.~Bahcall and R.K.~Ulrich, {\it Rev. Mod. Phys.} {\bf 60} (1989) 297.
\bibitem{L}
    A.~Lubicic et al., {\it Particle World} {\bf 2} (1991) 101.
\bibitem{MSW}
    L.~Wolfenstein, {\it Phys. Rev.} {\bf D17} (1977) 2369;
    S.P.~Mikheyev and A.Yu.~Smirnov, {\it Sov. J. Nucl. Phys.}
    {\bf 42} (1986) 913.
\bibitem{AV}
    T.M.~Aliyev and M.I.~Vysotsky, {\it Uspechi Fiz. Nauk} {\bf 135}
    (1981) 709 (in Russian).
\bibitem{ZMB}
    A.V.~Borisov et al., {\it Yad. Fiz.} {\bf 41},
    (1985) 743 ({\it Sov. J. Nucl. Phys.}, in Russian).
\bibitem{GMV}
    A.A.~Gvozdev, N.V.~Mikheev and L.A.~Vassilevskaya, {\it Phys. Lett.}
    {\bf B289} (1992) 103.
\bibitem{BTV}
    A.V.~Borisov, I.M.~Ternov and L.A.~Vassilevskaya, {\it Phys. Lett.}
    {\bf B273} (1991) 163.
\bibitem{KuM}
    A.V.~Kuznetsov and N.V.~Mikheev, {\it Phys. Lett.} {\bf B299} (1993) 367.
\bibitem{BLP}
    V.B.~Berestetskii, E.M.~Lifshitz and L.P.~Pitaevskii, {\it Quantum
    electrodynamics}, 2nd Ed. (Pergamon Press, Oxford, 1982).
\bibitem{S}
    J.~Schwinger, {\it Phys. Rev.} {\bf 82} (1951) 664.
\end{thebibliography}
\end{document}